\documentclass[apj]{emulateapj}
\usepackage{lscape,apjfonts}
\newcommand{\hii}         {\mbox{\rm \ion{H}{2}}}
\newcommand{\oiii}        {[\mbox{\rm \ion{O}{3}}] $\lambda$4363~\AA}

\def\spose#1{\hbox to 0pt{#1\hss}}
\def\lta{\mathrel{\spose{\lower 3pt\hbox{$\mathchar"218$}}
     \raise 2.0pt\hbox{$\mathchar"13C$}}}
\def\gta{\mathrel{\spose{\lower 3pt\hbox{$\mathchar"218$}}
     \raise 2.0pt\hbox{$\mathchar"13E$}}}

\shorttitle{Resolving Abundance Gradient Discrepancies in M33}
\shortauthors{Rosolowsky \& Simon}

\begin{document}

\title{The M33 Metallicity Project: Resolving the Abundance Gradient
  Discrepancies in M33\altaffilmark{1}}

\author{Erik Rosolowsky} 

\affil{Harvard-Smithsonian Center for Astrophysics, 60 Garden St.,
  MS-66, Cambridge, MA 02138; erosolow@cfa.harvard.edu}

\and 

\author{Joshua D. Simon}

\affil{Department of Astronomy, 
       California Institute of Technology, 1200 E. California Blvd.,
       MS 105-24, Pasadena, CA  91125; jsimon@astro.caltech.edu}

\altaffiltext{1}{Data presented herein were obtained at the
  W. M. Keck Observatory, which is operated as a scientific
  partnership among the California Institute of Technology, the
  University of California, and the National Aeronautics and Space
  Administration. The Observatory was made possible by the generous
  financial support of the W. M. Keck Foundation.}

\begin{abstract}
We present a new determination of the metallicity gradient in M33,
based on Keck/LRIS measurements of oxygen abundances using the
temperature-sensitive emission line [\ion{O}{3}]~$\lambda$4363~\AA\ in
61 \hii\ regions.  These data approximately triple the sample of
direct oxygen abundances in M33.  We find a central abundance of 12 +
log(O/H) = $8.36 \pm 0.04$ and a slope of $-0.027 \pm
0.012$~dex~kpc$^{-1}$, in agreement with infrared measurements of the
neon abundance gradient but much shallower than most previous oxygen
gradient measurements.  There is substantial intrinsic scatter of
0.11~dex in the metallicity at any given radius in M33, which imposes
a fundamental limit on the accuracy of gradient measurements that rely
on small samples of objects.  We also show that the ionization state
of neon does not follow the ionization state of oxygen as is commonly
assumed, suggesting that neon abundance measurements from optical
emission lines require careful treatment of the ionization
corrections.
\end{abstract}


\section{INTRODUCTION}
\label{intro}

M33 was one of the first galaxies in which a radial abundance gradient
was recognized \citep{searle71}.  Beginning with \citet{smith75}, a
series of studies showed that \hii\ region abundances in M33 decline
by approximately an order of magnitude over the visible extent of the
galaxy \citep{ka81,vilchez88,garnett92}, with a simple exponential
providing an accurate fit to the gradient.  A typical value for the
gradient from this work is $-0.12 \pm 0.02$~dex~kpc$^{-1}$
\citep[][hereafter V88]{vilchez88}.  More recently, \citet[][hereafter
  C06]{crockett06} presented new metallicity measurements for a small
sample of \hii\ regions that indicated nearly constant oxygen
abundances across M33, with a best-fit gradient of only $-0.012 \pm
0.011$~dex~kpc$^{-1}$.  If confirmed, this very shallow gradient would
call into question a variety of results concerning, e.g., the
metallicity dependence of the Cepheid period-luminosity relation
\citep{lee02} and the CO-H$_{2}$ conversion factor \citep{eros03} that
depend on the abundance gradient in M33.  To complicate matters
further, infrared measurements of the neon abundance gradient, which
should be less sensitive to systematic effects such as extinction and
the temperature of the \hii\ region, yield an intermediate gradient of
$-0.034 \pm 0.015$~dex~kpc$^{-1}$ that is not in very good agreement
with any of the published oxygen gradients from optical data
\citep[][hereafter WNP02]{wnp02}.

In this paper, we present new, direct metallicity measurements based
on detections of the temperature-sensitive \oiii\ line in 61 \hii\
regions in M33, approximately tripling the sample available in the
literature for M33 (or any other galaxy).  We use these data to
provide an improved measurement of the abundance gradient in M33.  In
\S \ref{observations} we describe the observations, data reduction,
and derivation of metallicities.  In \S \ref{results} we determine the
oxygen abundance gradient in M33, attempt to derive neon abundances,
and discuss some of the implications of our results.  We briefly
summarize the paper in \S \ref{conclusions}.

\section{OBSERVATIONS AND DATA REDUCTION}
\label{observations}

\subsection{Observations}

The M33 Metallicity Project is a long-term program in which we plan to
obtain deep spectroscopy of $\sim1000$ \hii\ regions in M33 drawn from
the unified catalog of \citet{lgatlas}.  These data will eventually
yield a high-resolution, two-dimensional metallicity map covering the
entire galaxy, with a mean sampling rate of $\gta2$ measurements per
square kpc, and sampling down to $\lta50$~pc in some regions.  At
present, the data set is approximately 20\%\ of its expected final
size.

The observations discussed in this paper were obtained with the LRIS
spectrograph \citep{oke95} on the Keck I telescope on 2004 September
22-23 and 2004 October 21.  Weather conditions during the September
nights were clear, while the October observations were affected by
clouds at times.  The spectrograph was configured with the 600/4000
grism on the blue side, a dichroic at 5600~\AA, and the 900/5500 and
400/8500 gratings on the red side (in separate exposures).  The
blue-side spectral resolution was $\sim5$~\AA, and spectra typically
covered $3600-5400$~\AA\ with some variation resulting from slit
placement on the mask.  Slits for which the wavelength coverage missed
the [\ion{O}{2}]~$\lambda$3727~\AA\ line have been excluded from our
analysis.  The slits were 1\arcsec\ wide and had a minimum length of
10\arcsec.  We used integration times of 15 minutes per exposure, with
$1-4$ exposures per slitmask.  The slits were generally aligned within
$\sim10\degr$ of the parallactic angle during the observations.  We
observed 11 slitmasks in the southwest half of the galaxy, with an
average of $\sim20$ slits on each mask.  For the purposes of this
paper we use only the blue-side data, which include all of the lines
([\ion{O}{2}] $\lambda\lambda$3726, 3729\AA\ and
[\ion{O}{3}]$\lambda\lambda$ 4363, 4959, 5007 \AA) necessary to derive
oxygen abundances.

\subsection{Data Reduction}
\label{data-reduction}
The data were reduced using an IDL pipeline originally developed for
the Sloan Digital Sky Survey spectroscopic data (D. J. Schlegel et al., in
preparation) and then modified to deal appropriately with multi-slit
LRIS spectra.  The images were bias-subtracted and then flat-fielded
using either dome flat exposures or the internal quartz lamp.  We used
the flatfield exposures to identify the location of each slit on the
CCD, and the remainder of the processing was carried out on each slit
individually.  We produced a two-dimensional wavelength solution for
each slit using exposures of Hg, Cd, Zn, Ne, and Ar arc lamps.  We
then removed cosmic rays from the images and coadded the frames
together.

\begin{figure*}
\plotone{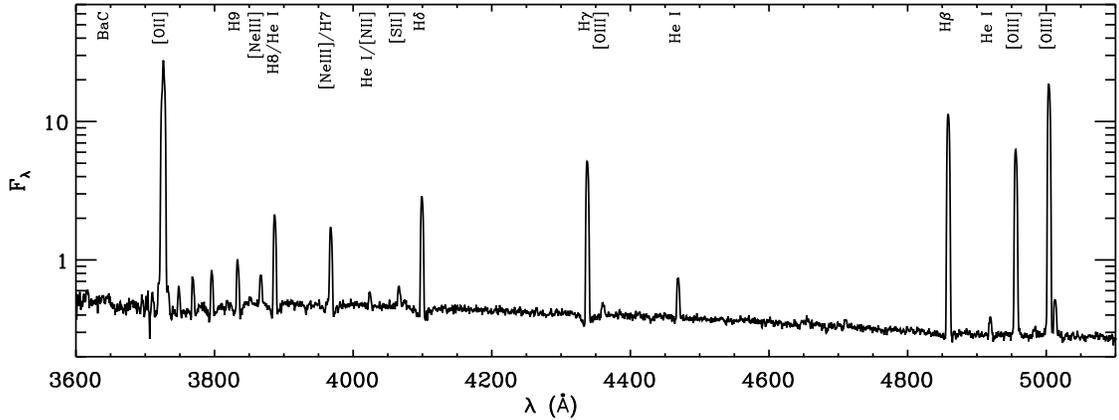}
\caption{\label{examplespec} Example spectrum of \hii\ region 60 in
  Table \ref{linetable} (BCLMP 266).  This spectrum has the median S/N
  (8.4) for the [\ion{O}{3}]~$\lambda$4363~\AA\ line in our sample.
  $F_{\lambda}$ is given in units of
  $10^{-16}$~erg~s$^{-1}$~cm$^{-2}$~\AA$^{-1}$.}
\end{figure*}

After these basic reduction steps, each slit was examined
interactively and the targets were identified based on peaks in the
spatial profile of the H$\beta$ line.  We selected an extraction
region around each target by following the H$\beta$ profile down to
the background level (which in most cases includes emission from the
diffuse ionized medium of M33) and setting the extraction limits where
the H$\beta$ intensity reached the background value on either side of
the peak.  A b-spline fit to the remainder of the slit produced a 2D
sky model covering the entire slit.  Finally, we subtracted the sky
model from the data and extracted the \hii\ region spectrum with a
boxcar algorithm, yielding a 1D spectrum and associated uncertainties.
The spectrum was corrected for atmospheric extinction and
flux-calibrated by comparison to an exposure of the spectrophotometric
standard star G191B2B \citep{massey88}.  An example calibrated 1D
spectrum is shown in Figure \ref{examplespec}.

Our procedure for measuring emission line fluxes in the 1D spectra
began by dividing each spectrum up into a series of short
($\sim200$~\AA) spectral windows, each containing at least one bright
line (except for the regions between 4500~\AA\ and 4800~\AA\ and
beyond 5100~\AA\ where no such lines exist) and performing a
constrained linear+Gaussian fit to all of the emission lines contained
within each window.  In each window the line separations were fixed at
their known values, and the widths of the lines were all set equal to
the width of the brightest line in the window (except for blends such
as the partially resolved [\ion{O}{2}]~$\lambda$3727~\AA\ doublet).
We then measured fluxes by integrating the observed intensity over a
$\pm1.5$ FWHM region centered on each line.  Uncertainties in the
measured line fluxes were calculated by propagating the uncertainty
associated with each pixel in the one-dimensional spectrum.  We
corrected the measured line fluxes for reddening and Balmer absorption
from the underlying stellar population using the method outlined by
\citet{os01}, including all of the Balmer emission lines for which we
had high S/N detections.  We used the Galactic reddening law of
\citet{ccm89} and assumed that the equivalent widths of all of the
Balmer absorption lines in a given \hii\ region were the same.  We
experimented with setting the ratios of the equivalent widths of the
absorption lines equal to their mean values averaged over time for the
solar metallicity, continuous star formation models of \citet{g-d99},
but we found that the resulting metallicity changes are always less
than 0.01~dex.  With such minimal changes and with possible
differences between the stellar populations simulated by \citet{g-d99}
and those present in M33 \hii\ regions, we proceeded using the simpler
model of constant Balmer equivalent widths.  Corrected line fluxes for
each \hii\ region with a detection of
[\ion{O}{3}]~$\lambda$4363~\AA\ are given in Table \ref{linetable}.
The listed uncertainties in the line fluxes have been propagated
through the reddening and absorption corrections.  Balmer line ratios
as well as derived reddening and absorption values are given in Table
\ref{hlinetable}.


\subsection{Derivation of \hii\ Region Temperatures and Abundances}
We measure oxygen abundances using the {\it nebular} module
\citep{sd95} of the STSDAS package in IRAF.  We model the nebula as
consisting of two ionization zones with the low- and medium-ionization
zones defined by the presence of O$^{+}$ and O$^{+2}$, respectively.
The temperature of the medium ionization zone is calculated using {\it
  nebular} based on the standard ratio of the
[\ion{O}{3}]~$\lambda\lambda 4959,5007$~\AA~doublet to the
[\ion{O}{3}]~$\lambda$4363~\AA\ auroral line.  To determine the
temperature of the low-ionization zone, we adopt the scaling of
\citet{campbell86}: $T([\mbox{\ion{O}{2}}]) =
0.70~T([\mbox{\ion{O}{3}}])+3000~\mathrm{K}.$ We assume that all
\ion{H}{2} regions are in the low density limit ($n_e \lta
10^2~\mbox{cm}^{-3}$).  With these densities and temperatures, we
measure ionic abundances relative to H$^+$ for O$^{+}$ and O$^{+2}$
and assume that all oxygen is found in these two ionization states.
This analytic approach is common to many extragalactic abundance
studies \citep[e.g.,][]{ken_m101}, but relies on several (admittedly
well-justified) assumptions.  We will examine these assumptions in
future work with the complete sample of \ion{H}{2} regions including
the full spectral coverage of LRIS.  

We restrict our analysis to the 61 \ion{H}{2} regions that have
detections of the [\ion{O}{3}]~$\lambda$4363~\AA\ line with S/N$ > 3$
(more conservative selections do not change our results; see \S
\ref{grad-sec}).  The locations of these \hii\ regions are
superimposed on an H$\alpha$ image of the galaxy in Figure
\ref{slitloc}.  All of our targets from this phase of the M33
Metallicity Project are in the southwest portion of the galaxy.
Previous work \citep[with the notable exception of][]{magrini07} has
focused primarily on giant \ion{H}{2} complexes (e.g., NGC~604,
NGC~588) in the northeast half of the galaxy.  The
[\ion{O}{3}]~$\lambda$4363~\AA\ line is detected in $\sim 1/3$ of our
targets and this fraction shows no significant variation with
galactocentric radius.  Contaminating supernova remnants, emission
line stars and planetary nebulae are removed from the sample based on
identification of characteristic spectral line features and
cross-matching with published catalogs of these objects.

As noted in \S\ref{data-reduction}, uncertainties in the line fluxes
are propagated from single pixel error values.  The uncertainties in
$c(\mathrm{H}\beta)$ and the equivalent width of the Balmer absorption
have been included in the errors for the dereddened line fluxes.  The
reported errors in the physical parameters are determined by a Monte
Carlo propagation of errors through the entire metallicity
determination process.  The Monte Carlo process is particularly
well-suited for gauging the widths of skewed uncertainty distributions
in the derived properties from low signal-to-noise lines.  For each
\ion{H}{2} region, we generate 100 sets of trial line fluxes by adding
a normal deviate times the flux uncertainty to each line flux and
re-analyzing the spectrum using {\it nebular}.  We also include a
random fluctuation in the temperature of the low ionization zone
derived from the relationship of \citet{campbell86}.  Based on the
magnitude of the uncertainties found in the work of \citet{ken_m101},
the 1$\sigma$ error in $T$([\ion{O}{2}]) is 300~K.  The final
uncertainties on the abundances are defined as the 1$\sigma$ widths of
the Monte Carlo abundance distributions (see also C06).  For
reasonably strong [\ion{O}{3}]~$\lambda$4363~\AA~emission (S/N$ >
10$), precisions in the derived metallicity are $\lta 0.08$ dex.  For
the four \ion{H}{2} regions in our sample that have been studied by
previous observers, the abundance determinations agree to within the
uncertainties (see Table \ref{comparison_table}).

\begin{deluxetable*}{lccccl}
\tablewidth{0pt}
\tablecolumns{6}
\tablecaption{Comparison with Literature Metallicity Measurements}
\tablehead{
\colhead{\hii\ region} & \colhead{KA81}  & 
\colhead{V88} & \colhead{C06} & \colhead{M07} &
\colhead{Our} \\
\colhead{} & \colhead{Abundance} & \colhead{Abundance} &
\colhead{Abundance} & \colhead{Abundance} & \colhead{Abundance}}
\startdata 
NGC~588   & \nodata & $8.30 \pm 0.06$ & \nodata       & \nodata & $8.315 \pm 0.061$  \\
MA~2      & 8.38    & $8.44 \pm 0.15$ & \nodata       & \nodata & $8.334 \pm 0.083$  \\
BCLMP~90  & \nodata &\nodata          & $8.50 \pm 0.06$ & \nodata & $8.503 \pm 0.057$\tablenotemark{a} \\
BCLMP~218 & \nodata &\nodata          & \nodata       & $8.25 \pm
0.05$ & $8.157 \pm 0.059$
\enddata
\tablerefs{KA81 = \citealt{ka81}; V88 = \citealt{vilchez88}; C06 =
  \citealt{crockett06}; M07 = \citealt{magrini07}}
\tablenotetext{a}{Weighted average of two independent measurements at
  the same position; see Table \ref{linetable}.}
\label{comparison_table}
\end{deluxetable*}

\subsection{\ion{He}{2} Zones}\label{heii}
We detect \ion{He}{2} $\lambda 4686$~\AA\ from 5 of the 61 \ion{H}{2}
regions in our sample at a typical level of 2.5\% of the H$\beta$ flux
(objects 3, 8, 16, 18, and 29 in Table \ref{linetable}).  The emission
is almost certainly nebular as it is spectrally unresolved but
spatially resolved in these sources (C06).  The presence of
\ion{He}{2} emission indicates that the simple two ionization zone
model adopted in the metallicity estimates is somewhat inaccurate.  A
high ionization region implies the presence of O$^{+3}$, which is not
accounted for in our abundance determination.  Consequently, the
derived oxygen abundances will be systematically low by an unknown
amount.  To estimate the magnitude of this effect, we calculate
He$^{+2}$/He$^{+}$ based on the ratio of the \ion{He}{2} $\lambda
4686$~\AA\ and \ion{He}{1} $\lambda 4471$~\AA\ lines.  This ratio is
only approximate because the \ion{He}{1} line is likely contaminated
by stellar absorption.  The derived ratio is roughly equal to
O$^{3+}$/(O$^{+}$+O$^{2+}$), providing an estimate of the contribution
of O$^{3+}$.  We therefore expect that we have underestimated the
oxygen abundance by $\sim 0.01-0.06$ dex in these regions.  Since we
are not sure of the influence of stellar absorption lines, we add (in
quadrature) this derived increase to the reported oxygen uncertainty
for these five regions.

\section{RESULTS}
\label{results}

\subsection{The Oxygen Gradient}
\label{grad-sec}
\begin{figure}
\plotone{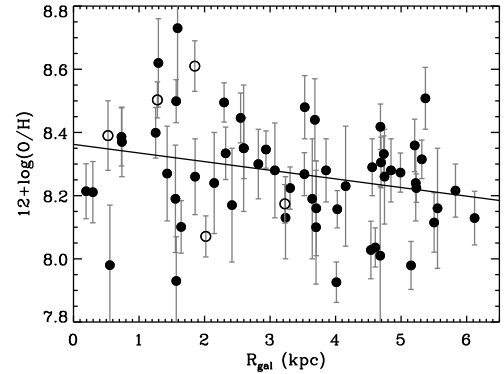}
\caption{\label{gradient} Abundances of 61 \ion{H}{2} regions in M33
  as a function of galactocentric radius.  A linear gradient with a
  slope of $-0.027$~dex~kpc$^{-1}$ is fit to the data (solid
  line).  Regions with significant \ion{He}{2}~$\lambda
  4686$\AA\ emission are
  indicated with open symbols.}
\end{figure}

Figure \ref{gradient} shows the radial distribution of gas phase
oxygen abundances in M33.  We have assumed a thin disk geometry based
on a distance to M33 of 840 kpc \citep{Freedman2001}, $i=52^{\circ}$
and PA = $22^{\circ}$ \citep{cs00}.  An ordinary, linear least-squares
fit to the data, weighting each point according to its inverse
variance, produces a gradient of $12+\log(\mathrm{O/H}) = (8.38 \pm
0.03) - (0.032 \pm 0.006)~R_{\rm kpc}$, but the residuals from this
fit are much larger than would be expected from the measurement
uncertainties ($\widetilde{\chi^{2}} = 3.2$).  Since it is evident
that an exponential decline with radius is a reasonable functional
form to describe the data, the high $\widetilde{\chi^{2}}$ value
indicates the presence of substantial intrinsic scatter around the
relation.

In this situation, accurately estimating the fit parameters must be
done using the method of \citet[][hereafter AB96]{bces}.  We perform a
linear regression with the AB96 technique for weighted least-squares
fitting in the presence of intrinsic scatter in the data.  The AB96
technique estimates the intrinsic variance and adds this variance to
that of the data to produce the appropriate weights for the fit.
Since the intrinsic variance is larger than the statistical
uncertainties for the M33 abundance data, there is less variation in
the weights than in the measurement uncertainties.  Even though
[\ion{O}{3}] $\lambda4363$~\AA\ detections with S/N as low as 3 result
in relatively large uncertainties in the derived abundances, including
these data improves the quality of the fit by increasing the sample
size.  This technique yields an abundance gradient of
\begin{equation}
12+\log(\mathrm{O/H}) = (8.36 \pm 0.04) - (0.027 \pm 0.012)~R_{kpc}.
\end{equation}

\begin{figure*}
\plotone{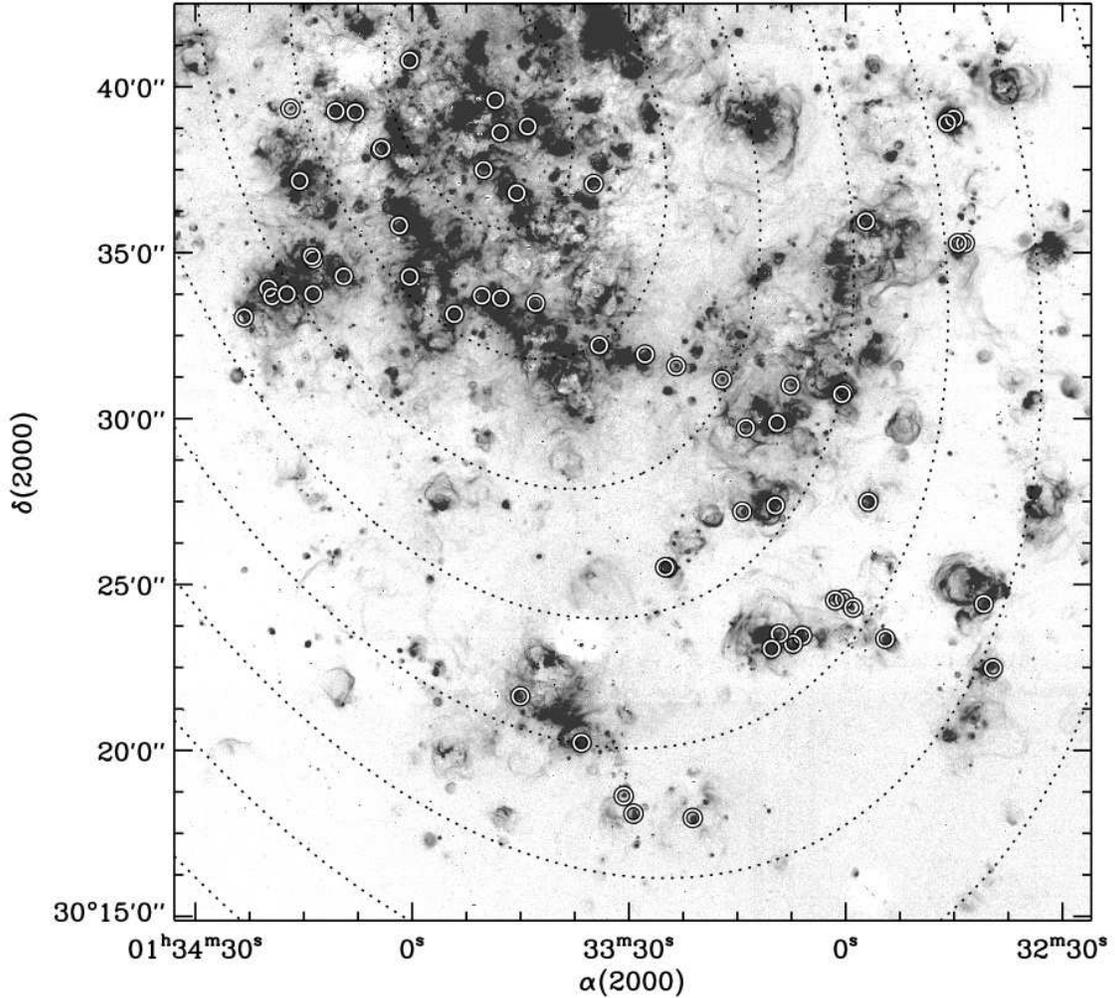}
\caption{\label{slitloc} Locations of the 61 \ion{H}{2} regions with
  oxygen abundance determinations superimposed on a
  continuum-subtracted H$\alpha$ image of the southwest portion of
  M33.  The image is produced from the survey data of
  \citet{massey-m33m31}.  Contours of constant galactocentric radius
  are shown as dotted ellipses with a spacing of 1 kpc.}
\end{figure*}

The implied intrinsic variance in the population (i.e., the scatter in
the metallicities at any given radius) is 0.11 dex, larger than the
precision of most of the measurements.  We check the uncertainties on
the central abundance and the slope of the abundance gradient by
bootstrapping the data \citep{numrec}, which yields uncertainties
$\lta 10\%$ higher than those derived from the AB96 formalism.
Although all the included data appear to have a statistically (and
visibly) significant [\ion{O}{3}] $\lambda 4363$~\AA\ detection,
performing the analysis only on the 23 \ion{H}{2} regions with
S/N$>10$ in this line yields an indistinguishable gradient of
$12+\log(\mathrm{O/H}) = (8.41 \pm 0.07) - (0.039 \pm 0.021)~R_{kpc}$.
We note that all of our targets are located in southwest half of the
galaxy (see Figure \ref{slitloc}), and therefore that the other half
of the galaxy could conceivably have a different gradient, but we
regard this possibility as unlikely.  Finally, we have correlated our
derived metallicities and their residuals around a radial gradient
against numerous parameters in our analysis such as L(H$\alpha$) taken
from the catalog of \citet{lgatlas}, $c$(H$\beta)$, the equivalent
widths of the Balmer absorption and H$\beta$ emission, and the oxygen
ionization correction factor.  We find no correlation with the
residuals that would indicate systematic errors in our methods.

\subsection{Complications in Measuring the Neon Gradient}
\begin{figure}
\plotone{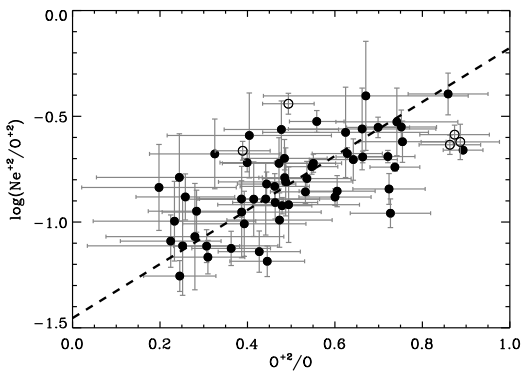}
\caption{\label{neon-problem} Ratio of doubly-ionized neon to
  doubly-ionized oxygen plotted as a function of O$^{+2}$ ionization
  fraction.  Based on photoionization models, this value should be
  constant and equal to the Ne/O ratio.  However, there is a clear
  trend of the ratio between the doubly ionized species and the degree
  of ionization in real \ion{H}{2} regions.  The dashed line shows the
  fit given in Equation \ref{neonfit}. Regions with significant
  \ion{He}{2}~$\lambda 4686$\AA\ emission are indicated with open
  symbols.}
\end{figure}

The slope of the neon abundance gradient is expected to be equal to
that of oxygen since these elements are produced in nearly equal
proportions for all channels of stellar nucleosynthesis.  WNP02
derived neon abundances from \emph{Infrared Space Observatory} spectra
and found a gradient consistent with our oxygen data.  Optical
observers commonly derive a neon gradient based on the
[\ion{Ne}{3}]~$\lambda$3869~\AA\ line, yielding an Ne$^{+2}$
abundance.  The fraction of neon found in the Ne$^{+2}$ state is often
assumed to be the same as the fraction of oxygen in O$^{+2}$, so that
an identical ionization correction factor (ICF) can be used for the
two elements \citep[C06]{stasinska01}.  Using the simple ICF results,
we found a significant gradient in the Ne/O ratio, contrary to
expectations.  We find cause for concern with this assumption for the
neon ICF when we derive the abundance of Ne$^{+2}$ using the standard
method and plot the Ne$^{+2}$ to O$^{+2}$ ratio as a function of the
fraction of oxygen found in the doubly ionized state (Figure
\ref{neon-problem}).  This ratio should be invariant with oxygen
ionization fraction and equal to the Ne/O ratio, yet the plot shows a
significant slope, with Ne$^{+2}$/O$^{+2}$ varying by a factor of
$\sim4$ over the observed range.  Since the relationship between the
two variables is not known a priori, we fit a linear regression using
the BCES method of AB96 to account for the uncertainties in both
directions.  The resulting fit gives:
\begin{equation}
\log(\mbox{Ne$^{+2}$/O$^{+2}$}) = (-1.45 \pm 0.10) + 
(1.27 \pm 0.22)~\mbox{O$^{+2}$/O}.
\label{neonfit}
\end{equation}
This analysis finds that the slope of the relationship is strongly
inconsistent with the value of zero expected from using the same ICF
for neon and oxygen.  The fraction of doubly-ionized oxygen is a proxy
for the ionization state of the nebula, so the trend indicates less
Ne$^{+2}$ relative to O$^{+2}$ in low ionization nebulae.  This is the
expected direction for this trend since the third ionization potential
of neon (41 eV) is larger than that of oxygen (35 eV).  More recent
work by \citet{izotov04,izotov06} suggests that applying the oxygen
ICF to neon may not be valid in lower ionization regions because of
charge transfer reactions between O$^{2+}$ and atomic hydrogen.  These
discrepancies could also be caused by a failure to account
appropriately for photons with $h\nu \sim 40$~eV in the O-star
atmospheres used in photoionization codes (C06).

\citet{perez-montero07} propose a nonlinear ICF based on their own
photoionization models:
\begin{equation}
\mathrm{ICF(Ne^{2+})}= 0.753+0.142x+\frac{0.171}{x},
\end{equation}
where $x=\mathrm{O^{2+}/(O^++O^{2+})}$.  Adopting this ICF, we
recalculate neon abundances and find no significant gradient in Ne/O
with radius.  For our sample of \hii\ regions, we measure
$\langle\mathrm{log(Ne/O)}\rangle=-0.69\pm 0.07$, consistent with the
results of \citet{perez-montero07} for giant extragalactic \ion{H}{2}
regions.  We conclude that using the same ICF for neon and oxygen is
inappropriate for the \ion{H}{2} regions we observed and that more
sophisticated ICFs seem to produce reasonable results for generic
\ion{H}{2} regions in M33.

\subsection{Comparison to Previous Gradient Measurements}

Our measured gradient of $-0.027$~dex~kpc$^{-1}$ is consistent within
the uncertainties with the neon gradient of WNP02 ($-0.034\pm 0.015$
dex kpc$^{-1}$) and marginally so with the oxygen gradient of C06
($-0.012\pm 0.011$ dex kpc$^{-1}$).  However, it is not consistent
with that of V88 ($-0.05\pm 0.01$ dex kpc$^{-1}$ for the outer regions
and $-0.10$ dex kpc$^{-1}$ overall after rescaling to our assumed
distance) and various older studies.  We note that the steep inner
gradient measured by V88 is driven by abundances determined from
photoionization models, which may be systematically high compared to
electron temperature abundances.  The apparent discrepancy in
abundance gradient likely results, at least in part, from an
underestimation of the uncertainties imparted by intrinsic scatter in
the \ion{H}{2} region abundances (see \S\ref{variance}).  The
intrinsic scatter in our optical \hii\ region abundances (0.11 dex) is
comparable to the value of $0.07-0.10$~dex reported by WNP02.

WNP02 suggested that a linear gradient may not be an appropriate
description for the enrichment of M33, arguing for a flat distribution
of abundances with a step down at a galactocentric radius of 4~kpc.
We fit a linear gradient to the 38 \ion{H}{2} regions inside $R_{\rm
  gal}=4$~kpc and find a slope of $-0.015 \pm 0.024$~dex~kpc$^{-1}$,
which is consistent with no gradient.  \citet{stan06-pne} also claimed
an abundance plateau in the inner galaxy based on several sub-solar
metallicity planetary nebulae in the region.  We measure a mean
metallicity in the inner galaxy ($R_{\rm gal}<4$~kpc) of 8.30; in the
outer galaxy, the mean is 8.21, implying a jump of the same order as,
but somewhat smaller than, that of WNP02 if a step function
metallicity profile holds.

There now appears to be a growing consensus that the oxygen abundance
gradient in \hii\ regions in M33 is relatively shallow, in conflict
with the decades-long assumption of a gradient at least as steep as
$-0.10$~dex~kpc$^{-1}$.  If we ignore the innermost \hii\ region
observed by V88, which does not have a detection of [\ion{O}{3}]
$\lambda 4363$~\AA\ or any other temperature-sensitive lines, then all
of the modern \ion{H}{2} region studies report gradients of
$-0.06$~dex~kpc$^{-1}$ or shallower \citep[V88, C06,][]{magrini07}.
We show in \S \ref{variance} that the results of these studies, which
include $6-14$ \hii\ region abundances, are consistent with our
measurements once their small sample sizes are taken into account.

Other types of sources besides \hii\ regions are increasingly being
employed to study abundances in M33.  \citet{urbaneja05} used
quantitative spectroscopy of OB supergiants to determine an oxygen
gradient of $-0.06 \pm 0.02$~dex~kpc$^{-1}$.  Since these stars are
very young, they are expected to trace the interstellar medium (ISM)
abundances.  If the OB stars have a similar intrinsic abundance
scatter to the \hii\ regions we measure, then this result would be
compatible with our findings.  \citet{magrini07} argue for a
steepening of the gradient at small radii based on the young star
metallicities in the inner 2 kpc of the galaxy \citep{urbaneja05}, but
we find no evidence for such a steepening based on the \ion{H}{2}
region data alone.  Planetary nebulae \citep{magrini04} still yield a
steep oxygen gradient ($-0.14$~dex~kpc$^{-1}$), but this result is
dependent on a single nebula at large radius; the remaining objects
clearly have a much shallower gradient.  Finally, supernova remnants
have also shown a shallow oxygen gradient, although the gradients in
other elements appear to be steeper \citep{bk85}.

\subsection{Intrinsic Scatter and the Determination of Gradients}
\label{variance}
Even after accounting for uncertainties in the stellar absorption and
reddening corrections, we find that there is an intrinsic scatter of
0.11 dex around the gradient that is unexplained by the measurement
uncertainties.  Such scatter is commonly seen in gradient
determinations \citep[e.g., in the Milky Way;][]{gal-abund-grad} and
may result from metallicity fluctuations in the interstellar medium.
Regardless of its source, gradient determinations made in the face of
significant scatter coupled with a limited number of observations may
produce widely varying results.  The historical evolution of the
gradient determination in M33 (one of the best-studied galaxies),
ranging over nearly an order of magnitude, serves as a cautionary
example.

\begin{figure}
\plotone{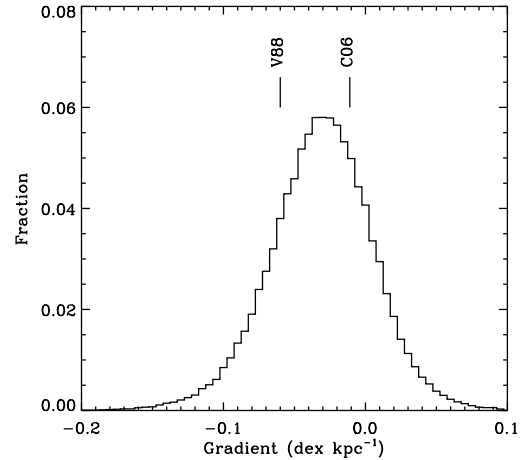}
\caption{\label{gradient-distn} Distribution of the abundance gradients
  that would be measured for samples of 10 \ion{H}{2} regions drawn
  randomly from our sample of 61.  The large width of distribution
  shows the uncertainty imprinted upon the abundance determinations by
  the underlying variance.  The measurements of \citet[V88]{vilchez88}
  and \citet[C06]{crockett06} are indicated and are consistent with
  the distribution given their small sample sizes. }
\end{figure}

Only large numbers of measurements can overcome the uncertainties
engendered by the intrinsic variance and relatively shallow gradient
in M33.  In Figure \ref{gradient-distn}, we show the likelihood of
measuring a particular value of the gradient if only 10 \hii\ regions
from our sample were used in the determination of the gradient.  We
model the behavior for 10 objects since this is a typical number of
targets studied in estimates of galactic abundance gradients, such as
those of V88 and C06.  The broad distribution of the derived gradients
(standard deviation of 0.03~dex~kpc$^{-1}$) suggests that
uncertainties in the gradients are systematically underreported in
galaxies with significant scatter around their gradients.  Even with a
sample several times larger than previous studies, we can only report
the presence of a gradient in M33 with 2.3$\sigma$ confidence!  Other
galaxies may have better established gradients owing to smaller
intrinsic scatter and/or a steeper abundance gradient \citep[e.g.,
M101;][]{ken_m101}.

\section{SUMMARY AND CONCLUSIONS}
\label{conclusions}

Using data from the M33 Metallicity Project, we have obtained direct
oxygen abundance measurements based on detections of the [\ion{O}{3}]
$\lambda 4363$~\AA\ line for 61 \ion{H}{2} regions in the southwest
half of M33.  This data set approximately triples the sample of M33
\hii\ region metallicities in the literature and makes M33 by far the
best-studied external galaxy in terms of its ISM abundances.  We
presented a refined oxygen gradient for M33 and found an exponential
abundance profile with a gradient of $-0.027 \pm 0.012$~dex~kpc$^{-1}$
and a central abundance of $12+\log(\mathrm{O/H})=8.36\pm 0.04$.
There is an intrinsic scatter of 0.11 dex around the trend, which
complicates the measurement of the shallow gradient in the galaxy.
When this scatter is accounted for, previous studies (which generally
derived steeper gradients) were shown to be consistent with our
measurements.  We also noted that there is a significant correlation
of the Ne$^{+2}$/O$^{+2}$ ratio with the fraction of oxygen found in
the doubly ionized state, which suggests that neon determinations
based solely on optical lines require more sophisticated ionization
correction factors than are typically assumed.

\acknowledgements{The authors wish to recognize and acknowledge the
  very significant cultural role and reverence that the summit of
  Mauna Kea has always had within the indigenous Hawaiian community.
  We are most fortunate to have the opportunity to conduct
  observations from this mountain.  We thank the anonymous referee for
  comments that improved the paper.  ER acknowledges support from an
  NSF AAP Fellowship (AST-0502605), and JDS acknowledges the support
  of a Millikan Fellowship provided by Caltech.  We thank Leo Blitz
  for support of the early stages of this project, Joe Hennawi for
  assistance with our data reduction code, and Jos{\'e} V{\' i}lchez
  for helpful conversations.  We also acknowledge the many useful
  discussions with the participants in the Metals 2007 Conference
  whose advice has improved this work.  This research has made use of
  NASA's Astrophysics Data System Bibliographic Services and the
  SIMBAD database, operated at CDS, Strasbourg, France.}

{\it Facilities:} \facility{Keck:I (LRIS)}


\clearpage
\LongTables
\begin{landscape}
\begin{deluxetable}{lcccccccccc}
\tabletypesize{\tiny}
\tablecaption{\label{linetable} \hii\ Region Line Fluxes}
\tablewidth{0pt}
\tablehead{ \colhead{Obj.} &
\colhead{Coordinates\tablenotemark{a}} & \colhead{Name\tablenotemark{b}} &\colhead{$R_{gal}$} &
\colhead{[\ion{O}{2}]\tablenotemark{c}} & 
\colhead{[\ion{Ne}{3}]\tablenotemark{c}} &
\colhead{[\ion{O}{3}]\tablenotemark{c}} &
\colhead{[\ion{O}{3}]\tablenotemark{c}} & 
\colhead{[\ion{O}{3}]\tablenotemark{c}} & 
\colhead{$T_{e}$([\ion{O}{3}])} &
\colhead{$12+\log([\mathrm{O/H}])$}\\
\colhead{} &
\colhead{($\alpha_{2000},\delta_{2000}$)} &
\colhead{} &
\colhead{(kpc)} &
\colhead{$\lambda 3727$~\AA} & 
\colhead{$\lambda 3869$~\AA} & 
\colhead{$\lambda 4363$~\AA} & 
\colhead{$\lambda 4959$~\AA} & 
\colhead{$\lambda 5007$~\AA} & 
\colhead{(K)} &
\colhead{}
}
\startdata
1 &  01:33:48.4  +30:39:36 & B0043b & 0.19 & $\phn1.42 \pm 0.01\phn$  & $\phn0.086 \pm 0.002\phn$  & $\phn\phn0.010 \pm 0.001\phn\phn$  & $\phn0.680 \pm 0.002\phn$  & $2.039 \pm 0.006$  & $\phn9200 \pm 300\phn\phn$  & $8.214 \pm 0.087$  \\
2 &  01:33:47.8  +30:38:38 & B0029 & 0.30 & $\phn1.58 \pm 0.01\phn$  & $0.0437 \pm 0.0007$  & $\phn0.0046 \pm 0.0005\phn$  & $\phn0.409 \pm 0.001\phn$  & $1.229 \pm 0.003$  & $\phn8600 \pm 300\phn\phn$  & $8.211 \pm 0.097$  \\
3 &  01:33:44.0  +30:38:49 & B0038b & 0.53 & $\phn0.85 \pm 0.03\phn$  & $\phn\phn0.51 \pm 0.02\phn\phn$  & $\phn\phn0.045 \pm 0.004\phn\phn$  & $\phn\phn2.25 \pm 0.03\phn\phn$  & $\phn6.50 \pm 0.07\phn$  & $10200 \pm 400\phn\phn$  & $8.391 \pm 0.090$  \\
4 &  01:33:50.1  +30:37:30 & B0016 & 0.56 & $\phn2.29 \pm 0.02\phn$  & $\phn0.023 \pm 0.001\phn$  & $\phn\phn0.005 \pm 0.001\phn\phn$  & $\phn0.231 \pm 0.001\phn$  & $0.694 \pm 0.003$  & $10300 \pm 900\phn\phn$  & $\phn7.98 \pm 0.19\phn$  \\
5 &  01:34:00.3  +30:40:48 & B0079c & 0.73 & $\phn1.26 \pm 0.05\phn$  & $\phn0.098 \pm 0.004\phn$  & $\phn\phn0.012 \pm 0.001\phn\phn$  & $\phn\phn1.02 \pm 0.01\phn\phn$  & $\phn3.05 \pm 0.03\phn$  & $\phn8700 \pm 200\phn\phn$  & $8.386 \pm 0.092$  \\
6 &  01:33:45.5  +30:36:49 & B0027b & 0.74 & $\phn2.75 \pm 0.01\phn$  & $0.0303 \pm 0.0006$  & $\phn0.0032 \pm 0.0004\phn$  & $0.3214 \pm 0.0008$  & $0.957 \pm 0.002$  & $\phn8400 \pm 300\phn\phn$  & $\phn8.37 \pm 0.11\phn$  \\
7 &  01:33:34.9  +30:37:05 & B0033b & 1.25 & $1.801 \pm 0.009$  & $0.0678 \pm 0.0007$  & $\phn0.0054 \pm 0.0004\phn$  & $\phn0.589 \pm 0.001\phn$  & $1.771 \pm 0.003$  & $\phn8200 \pm 200\phn\phn$  & $8.399 \pm 0.080$  \\
8\tablenotemark{d} &  01:34:04.2  +30:38:09 & B0090 & 1.28 & $\phn1.25 \pm 0.05\phn$  & $\phn\phn0.57 \pm 0.02\phn\phn$  & $\phn\phn0.049 \pm 0.003\phn\phn$  & $\phn\phn2.55 \pm 0.02\phn\phn$  & $\phn7.73 \pm 0.07\phn$  & $10000 \pm 200\phn\phn$  & $8.506 \pm 0.070$  \\
8\tablenotemark{d} &  01:34:04.2  +30:38:09 & B0090 & 1.28 & $\phn1.81 \pm 0.03\phn$  & $\phn0.652 \pm 0.009\phn$  & $\phn\phn0.048 \pm 0.002\phn\phn$  & $\phn\phn2.47 \pm 0.02\phn\phn$  & $\phn7.39 \pm 0.06\phn$  & $10100 \pm 100\phn\phn$  & $8.500 \pm 0.064$  \\
9 &  01:34:04.3  +30:38:07 & B0090 & 1.30 & $\phn\phn2.7 \pm 0.2\phn\phn$  & $\phn\phn0.69 \pm 0.05\phn\phn$  & $\phn\phn0.041 \pm 0.009\phn\phn$  & $\phn\phn2.56 \pm 0.07\phn\phn$  & $\phn\phn7.4 \pm 0.2\phn\phn$  & $\phn9600 \pm 500\phn\phn$  & $\phn8.62 \pm 0.14\phn$  \\
10 &  01:34:07.9  +30:39:14 & C204b & 1.43 & $\phn\phn2.7 \pm 0.1\phn\phn$  & $\phn0.072 \pm 0.004\phn$  & $\phn\phn0.010 \pm 0.003\phn\phn$  & $\phn0.602 \pm 0.007\phn$  & $\phn1.83 \pm 0.02\phn$  & $\phn9500 \pm 600\phn\phn$  & $\phn8.27 \pm 0.15\phn$  \\
11 &  01:33:47.7  +30:33:38 & B0018 & 1.56 & $\phn1.35 \pm 0.04\phn$  & $\phn0.220 \pm 0.007\phn$  & $\phn\phn0.014 \pm 0.004\phn\phn$  & $\phn\phn0.88 \pm 0.01\phn\phn$  & $\phn2.54 \pm 0.03\phn$  & $\phn9600 \pm 800\phn\phn$  & $\phn8.19 \pm 0.17\phn$  \\
12\tablenotemark{d} &  01:33:42.9  +30:33:30 & B0023 & 1.57 & $\phn1.63 \pm 0.03\phn$  & $\phn0.477 \pm 0.009\phn$  & $\phn\phn0.021 \pm 0.002\phn\phn$  & $\phn1.725 \pm 0.009\phn$  & $\phn4.93 \pm 0.02\phn$  & $\phn8800 \pm 300\phn\phn$  & $\phn8.55 \pm 0.11\phn$  \\
12\tablenotemark{d} &  01:33:42.9  +30:33:30 & B0023 & 1.57 & $\phn1.81 \pm 0.04\phn$  & $\phn\phn0.47 \pm 0.01\phn\phn$  & $\phn\phn0.031 \pm 0.004\phn\phn$  & $\phn\phn1.80 \pm 0.01\phn\phn$  & $\phn5.30 \pm 0.03\phn$  & $\phn9700 \pm 300\phn\phn$  & $8.444 \pm 0.081$  \\
13 &  01:34:01.8  +30:35:49 & B0001 & 1.57 & $\phn1.68 \pm 0.02\phn$  & $\phn0.035 \pm 0.002\phn$  & $\phn\phn0.008 \pm 0.002\phn\phn$  & $\phn0.379 \pm 0.003\phn$  & $1.071 \pm 0.005$  & $10400 \pm 700\phn\phn$  & $\phn7.93 \pm 0.14\phn$  \\
14 &  01:33:50.3  +30:33:42 & B0015b & 1.59 & $\phn\phn3.2 \pm 0.2\phn\phn$  & $\phn\phn0.68 \pm 0.05\phn\phn$  & $\phn\phn0.023 \pm 0.007\phn\phn$  & $\phn\phn2.01 \pm 0.04\phn\phn$  & $\phn\phn5.9 \pm 0.1\phn\phn$  & $\phn8600 \pm 800\phn\phn$  & $\phn8.73 \pm 0.23\phn$  \\
15 &  01:34:10.5  +30:39:16 & C208 & 1.64 & $\phn2.89 \pm 0.02\phn$  & $0.0167 \pm 0.0008$  & $\phn0.0063 \pm 0.0006\phn$  & $\phn0.294 \pm 0.001\phn$  & $0.947 \pm 0.002$  & $10200 \pm 300\phn\phn$  & $8.101 \pm 0.083$  \\
16 &  01:34:00.3  +30:34:17 & C001Ab & 1.85 & $\phn1.22 \pm 0.02\phn$  & $\phn\phn0.63 \pm 0.01\phn\phn$  & $\phn\phn0.041 \pm 0.003\phn\phn$  & $\phn\phn2.64 \pm 0.02\phn\phn$  & $\phn8.03 \pm 0.06\phn$  & $\phn9400 \pm 200\phn\phn$  & $8.610 \pm 0.080$  \\
17 &  01:33:54.1  +30:33:10 & B0013c & 1.86 & $\phn1.68 \pm 0.01\phn$  & $0.0394 \pm 0.0009$  & $\phn0.0049 \pm 0.0008\phn$  & $\phn0.451 \pm 0.002\phn$  & $1.349 \pm 0.004$  & $\phn8500 \pm 400\phn\phn$  & $\phn8.26 \pm 0.12\phn$  \\
18 &  01:33:34.1  +30:32:13 & B0208f & 2.02 & $\phn2.69 \pm 0.02\phn$  & $\phn0.127 \pm 0.002\phn$  & $\phn0.0147 \pm 0.0009\phn$  & $\phn0.580 \pm 0.002\phn$  & $1.742 \pm 0.005$  & $10900 \pm 200\phn\phn$  & $8.071 \pm 0.063$  \\
19 &  01:34:16.8  +30:39:21 & B0723 & 2.15 & $\phn\phn2.4 \pm 0.2\phn\phn$  & $\phn0.088 \pm 0.009\phn$  & $\phn\phn0.014 \pm 0.003\phn\phn$  & $\phn\phn0.78 \pm 0.02\phn\phn$  & $\phn2.30 \pm 0.05\phn$  & $\phn9800 \pm 700\phn\phn$  & $\phn8.24 \pm 0.16\phn$  \\
20 &  01:33:27.7  +30:31:56 & B0213b & 2.30 & $\phn4.10 \pm 0.06\phn$  & $\phn0.493 \pm 0.008\phn$  & $\phn\phn0.034 \pm 0.002\phn\phn$  & $\phn1.718 \pm 0.009\phn$  & $\phn5.13 \pm 0.03\phn$  & $10100 \pm 200\phn\phn$  & $8.495 \pm 0.062$  \\
21 &  01:34:15.6  +30:37:11 & MA2 & 2.32 & $1.332 \pm 0.005$  & $0.0626 \pm 0.0005$  & $\phn0.0053 \pm 0.0003\phn$  & $0.5695 \pm 0.0009$  & $1.733 \pm 0.002$  & $\phn8200 \pm 100\phn\phn$  & $8.334 \pm 0.083$  \\
22 &  01:34:09.5  +30:34:18 & C184A & 2.42 & $\phn2.31 \pm 0.04\phn$  & $\phn0.049 \pm 0.003\phn$  & $\phn\phn0.008 \pm 0.002\phn\phn$  & $\phn0.470 \pm 0.004\phn$  & $1.401 \pm 0.009$  & $\phn9600 \pm 800\phn\phn$  & $\phn8.17 \pm 0.18\phn$  \\
23 &  01:33:23.4  +30:31:35 & C129a & 2.55 & $\phn2.65 \pm 0.01\phn$  & $0.0859 \pm 0.0009$  & $\phn0.0074 \pm 0.0006\phn$  & $\phn0.682 \pm 0.002\phn$  & $2.038 \pm 0.005$  & $\phn8500 \pm 200\phn\phn$  & $8.446 \pm 0.079$  \\
24 &  01:34:13.8  +30:34:54 & B0714l & 2.60 & $\phn2.97 \pm 0.03\phn$  & $\phn0.026 \pm 0.002\phn$  & $\phn\phn0.004 \pm 0.001\phn\phn$  & $\phn0.348 \pm 0.002\phn$  & $1.042 \pm 0.005$  & $\phn8700 \pm 800\phn\phn$  & $\phn8.35 \pm 0.20\phn$  \\
25 &  01:34:13.6  +30:34:49 & B0714g & 2.60 & $\phn2.49 \pm 0.01\phn$  & $0.0522 \pm 0.0008$  & $\phn0.0053 \pm 0.0006\phn$  & $\phn0.475 \pm 0.001\phn$  & $1.409 \pm 0.003$  & $\phn8600 \pm 300\phn\phn$  & $8.350 \pm 0.096$  \\
26 &  01:34:13.7  +30:33:45 & S14 & 2.82 & $\phn3.20 \pm 0.02\phn$  & $\phn0.041 \pm 0.001\phn$  & $\phn0.0049 \pm 0.0007\phn$  & $\phn0.347 \pm 0.001\phn$  & $1.046 \pm 0.003$  & $\phn9100 \pm 400\phn\phn$  & $\phn8.30 \pm 0.11\phn$  \\
27 &  01:33:17.1  +30:31:11 & C121 & 2.94 & $\phn2.27 \pm 0.01\phn$  & $\phn0.094 \pm 0.001\phn$  & $\phn0.0082 \pm 0.0005\phn$  & $\phn0.661 \pm 0.002\phn$  & $1.955 \pm 0.005$  & $\phn8900 \pm 100\phn\phn$  & $8.346 \pm 0.059$  \\
28 &  01:34:17.3  +30:33:46 & B0711a & 3.07 & $\phn2.54 \pm 0.02\phn$  & $\phn0.069 \pm 0.001\phn$  & $\phn0.0046 \pm 0.0009\phn$  & $\phn0.377 \pm 0.002\phn$  & $1.118 \pm 0.004$  & $\phn8800 \pm 600\phn\phn$  & $\phn8.28 \pm 0.15\phn$  \\
29 &  01:34:19.9  +30:33:56 & C192 & 3.23 & $\phn4.18 \pm 0.07\phn$  & $\phn\phn0.56 \pm 0.01\phn\phn$  & $\phn\phn0.053 \pm 0.003\phn\phn$  & $\phn\phn1.44 \pm 0.01\phn\phn$  & $\phn4.25 \pm 0.03\phn$  & $12500 \pm 300\phn\phn$  & $8.174 \pm 0.059$  \\
30 &  01:34:19.3  +30:33:41 & B0712a & 3.24 & $\phn\phn3.5 \pm 0.2\phn\phn$  & $\phn\phn0.33 \pm 0.02\phn\phn$  & $\phn\phn0.038 \pm 0.007\phn\phn$  & $\phn\phn1.11 \pm 0.03\phn\phn$  & $\phn3.38 \pm 0.08\phn$  & $12100 \pm 800\phn\phn$  & $\phn8.13 \pm 0.13\phn$  \\
31 &  01:33:13.9  +30:29:44 & C023 & 3.31 & $\phn2.44 \pm 0.02\phn$  & $\phn0.090 \pm 0.002\phn$  & $\phn\phn0.011 \pm 0.001\phn\phn$  & $\phn0.635 \pm 0.002\phn$  & $1.885 \pm 0.006$  & $\phn9700 \pm 200\phn\phn$  & $8.224 \pm 0.067$  \\
32 &  01:33:07.6  +30:31:01 & C403 & 3.52 & $\phn2.26 \pm 0.02\phn$  & $0.0720 \pm 0.0009$  & $0.01000 \pm 0.0006\phn$  & $\phn0.661 \pm 0.002\phn$  & $1.970 \pm 0.005$  & $\phn9300 \pm 100\phn\phn$  & $8.268 \pm 0.068$  \\
33 &  01:33:09.5  +30:29:52 & B0217a & 3.53 & $\phn2.35 \pm 0.03\phn$  & $\phn0.108 \pm 0.002\phn$  & $\phn\phn0.008 \pm 0.001\phn\phn$  & $\phn0.783 \pm 0.004\phn$  & $\phn2.41 \pm 0.01\phn$  & $\phn8400 \pm 200\phn\phn$  & $\phn8.48 \pm 0.10\phn$  \\
34 &  01:34:23.2  +30:33:03 & B0715b & 3.64 & $\phn\phn4.2 \pm 0.2\phn\phn$  & $\phn\phn0.27 \pm 0.01\phn\phn$  & $\phn\phn0.031 \pm 0.008\phn\phn$  & $\phn\phn0.95 \pm 0.02\phn\phn$  & $\phn3.00 \pm 0.05\phn$  & $12000 \pm 1000\phn$  & $\phn8.19 \pm 0.19\phn$  \\
35 &  01:33:14.3  +30:27:11 & C034 & 3.68 & $\phn2.64 \pm 0.03\phn$  & $\phn0.116 \pm 0.003\phn$  & $\phn\phn0.008 \pm 0.002\phn\phn$  & $\phn0.719 \pm 0.003\phn$  & $2.122 \pm 0.008$  & $\phn8600 \pm 400\phn\phn$  & $\phn8.44 \pm 0.13\phn$  \\
36 &  01:33:25.0  +30:25:31 & B0222 & 3.70 & $\phn\phn3.5 \pm 0.1\phn\phn$  & $\phn0.042 \pm 0.003\phn$  & $\phn\phn0.007 \pm 0.002\phn\phn$  & $\phn0.301 \pm 0.003\phn$  & $0.854 \pm 0.008$  & $10700 \pm 900\phn\phn$  & $\phn8.10 \pm 0.18\phn$  \\
37 &  01:33:24.7  +30:25:32 & B0222 & 3.70 & $\phn1.98 \pm 0.03\phn$  & $\phn0.060 \pm 0.003\phn$  & $\phn\phn0.008 \pm 0.002\phn\phn$  & $\phn0.514 \pm 0.003\phn$  & $1.490 \pm 0.008$  & $\phn9500 \pm 700\phn\phn$  & $\phn8.16 \pm 0.15\phn$  \\
38 &  01:33:09.8  +30:27:23 & B0221 & 3.85 & $\phn1.59 \pm 0.04\phn$  & $\phn0.163 \pm 0.005\phn$  & $\phn\phn0.014 \pm 0.002\phn\phn$  & $\phn0.896 \pm 0.009\phn$  & $\phn2.71 \pm 0.03\phn$  & $\phn9300 \pm 400\phn\phn$  & $\phn8.28 \pm 0.10\phn$  \\
39 &  01:33:00.5  +30:30:44 & B0218 & 4.01 & $\phn1.71 \pm 0.02\phn$  & $\phn0.144 \pm 0.003\phn$  & $\phn\phn0.022 \pm 0.001\phn\phn$  & $\phn0.740 \pm 0.004\phn$  & $\phn2.15 \pm 0.01\phn$  & $11700 \pm 300\phn\phn$  & $7.926 \pm 0.064$  \\
40 &  01:33:00.3  +30:30:47 & B0218 & 4.03 & $\phn1.97 \pm 0.02\phn$  & $\phn0.139 \pm 0.002\phn$  & $\phn0.0159 \pm 0.0008\phn$  & $\phn0.784 \pm 0.003\phn$  & $2.356 \pm 0.007$  & $10200 \pm 200\phn\phn$  & $8.157 \pm 0.059$  \\
41 &  01:32:57.2  +30:35:57 & B0233c & 4.15 & $\phn\phn4.8 \pm 0.1\phn\phn$  & $\phn0.087 \pm 0.006\phn$  & $\phn\phn0.013 \pm 0.004\phn\phn$  & $\phn0.511 \pm 0.007\phn$  & $\phn1.61 \pm 0.02\phn$  & $11000 \pm 1000\phn$  & $\phn8.23 \pm 0.19\phn$  \\
42 &  01:32:56.9  +30:27:29 & B0243 & 4.54 & $\phn\phn1.3 \pm 0.1\phn\phn$  & $\phn\phn\phn1.3 \pm 0.1\phn\phn\phn$  & $\phn\phn\phn0.15 \pm 0.02\phn\phn\phn$  & $\phn\phn\phn2.8 \pm 0.1\phn\phn\phn$  & $\phn\phn8.0 \pm 0.3\phn\phn$  & $14600 \pm 700\phn\phn$  & $8.028 \pm 0.091$  \\
43 &  01:33:09.2  +30:23:30 & B0257 & 4.56 & $\phn3.31 \pm 0.04\phn$  & $\phn0.043 \pm 0.002\phn$  & $\phn\phn0.011 \pm 0.001\phn\phn$  & $\phn0.610 \pm 0.003\phn$  & $1.837 \pm 0.009$  & $\phn9800 \pm 300\phn\phn$  & $8.290 \pm 0.090$  \\
44 &  01:33:10.3  +30:23:05 & B0255a & 4.61 & $\phn3.04 \pm 0.06\phn$  & $\phn0.140 \pm 0.003\phn$  & $\phn\phn0.022 \pm 0.001\phn\phn$  & $\phn0.707 \pm 0.003\phn$  & $2.088 \pm 0.009$  & $11800 \pm 200\phn\phn$  & $8.036 \pm 0.062$  \\
45 &  01:33:07.4  +30:23:13 & B0259a & 4.68 & $\phn3.01 \pm 0.07\phn$  & $\phn0.027 \pm 0.004\phn$  & $\phn\phn0.009 \pm 0.003\phn\phn$  & $\phn0.335 \pm 0.004\phn$  & $1.045 \pm 0.009$  & $11000 \pm 1000\phn$  & $\phn8.01 \pm 0.22\phn$  \\
46 &  01:33:06.1  +30:23:27 & B0258b & 4.68 & $\phn2.27 \pm 0.03\phn$  & $\phn0.464 \pm 0.007\phn$  & $\phn\phn0.033 \pm 0.002\phn\phn$  & $\phn\phn1.74 \pm 0.01\phn\phn$  & $\phn5.17 \pm 0.03\phn$  & $10000 \pm 200\phn\phn$  & $8.418 \pm 0.072$  \\
47 &  01:33:01.6  +30:24:31 & C022 & 4.69 & $\phn1.89 \pm 0.01\phn$  & $\phn0.202 \pm 0.002\phn$  & $\phn0.0171 \pm 0.0006\phn$  & $\phn1.037 \pm 0.002\phn$  & $3.081 \pm 0.007$  & $\phn9600 \pm 100\phn\phn$  & $8.305 \pm 0.080$  \\
48 &  01:33:45.0  +30:21:38 & B0248 & 4.74 & $1.300 \pm 0.007$  & $\phn0.191 \pm 0.001\phn$  & $\phn0.0172 \pm 0.0005\phn$  & $\phn1.154 \pm 0.002\phn$  & $3.454 \pm 0.007$  & $\phn9300 \pm 80\phn\phn\phn$  & $8.332 \pm 0.057$  \\
49 &  01:33:00.3  +30:24:33 & C021 & 4.75 & $\phn\phn2.7 \pm 0.1\phn\phn$  & $\phn0.126 \pm 0.007\phn$  & $\phn\phn0.016 \pm 0.003\phn\phn$  & $\phn\phn0.86 \pm 0.01\phn\phn$  & $\phn2.52 \pm 0.03\phn$  & $10000 \pm 600\phn\phn$  & $\phn8.26 \pm 0.15\phn$  \\
50 &  01:32:59.1  +30:24:18 & C018 & 4.85 & $\phn1.86 \pm 0.08\phn$  & $\phn\phn0.48 \pm 0.02\phn\phn$  & $\phn\phn0.052 \pm 0.007\phn\phn$  & $\phn\phn1.86 \pm 0.03\phn\phn$  & $\phn\phn5.9 \pm 0.1\phn\phn$  & $11100 \pm 500\phn\phn$  & $\phn8.28 \pm 0.10\phn$  \\
51 &  01:33:36.6  +30:20:14 & MA19 & 4.99 & $\phn1.40 \pm 0.01\phn$  & $\phn0.227 \pm 0.002\phn$  & $\phn0.0209 \pm 0.0008\phn$  & $\phn1.176 \pm 0.003\phn$  & $3.531 \pm 0.008$  & $\phn9800 \pm 100\phn\phn$  & $8.273 \pm 0.062$  \\
52 &  01:32:44.4  +30:35:17 & C400 & 5.15 & $\phn2.78 \pm 0.03\phn$  & $\phn0.033 \pm 0.002\phn$  & $\phn\phn0.012 \pm 0.001\phn\phn$  & $\phn0.426 \pm 0.002\phn$  & $1.257 \pm 0.006$  & $11400 \pm 400\phn\phn$  & $7.979 \pm 0.076$  \\
53 &  01:32:54.6  +30:23:22 & B0261 & 5.21 & $\phn1.03 \pm 0.01\phn$  & $\phn0.095 \pm 0.002\phn$  & $\phn0.0078 \pm 0.0009\phn$  & $\phn0.810 \pm 0.004\phn$  & $\phn2.35 \pm 0.01\phn$  & $\phn8300 \pm 200\phn\phn$  & $8.359 \pm 0.083$  \\
54 &  01:32:43.5  +30:35:17 & C398 & 5.22 & $\phn4.36 \pm 0.07\phn$  & $\phn0.034 \pm 0.002\phn$  & $\phn\phn0.009 \pm 0.002\phn\phn$  & $\phn0.421 \pm 0.003\phn$  & $1.263 \pm 0.007$  & $10400 \pm 500\phn\phn$  & $\phn8.24 \pm 0.12\phn$  \\
55 &  01:32:45.9  +30:38:54 & NGC588 & 5.23 & $0.670 \pm 0.003$  & $\phn0.419 \pm 0.002\phn$  & $\phn0.0481 \pm 0.0003\phn$  & $\phn1.925 \pm 0.003\phn$  & $5.663 \pm 0.008$  & $10930 \pm 20\phn\phn\phn$  & $8.224 \pm 0.045$  \\
56 &  01:32:45.0  +30:39:02 & NGC588 & 5.32 & $\phn1.78 \pm 0.02\phn$  & $\phn0.215 \pm 0.003\phn$  & $\phn0.0191 \pm 0.0007\phn$  & $\phn\phn1.17 \pm 0.01\phn\phn$  & $\phn3.36 \pm 0.03\phn$  & $\phn9600 \pm 100\phn\phn$  & $8.315 \pm 0.061$  \\
57 &  01:33:30.8  +30:18:38 & B0252 & 5.37 & $\phn3.00 \pm 0.03\phn$  & $\phn0.146 \pm 0.002\phn$  & $\phn\phn0.009 \pm 0.001\phn\phn$  & $\phn0.851 \pm 0.003\phn$  & $2.536 \pm 0.008$  & $\phn8600 \pm 300\phn\phn$  & $8.508 \pm 0.098$  \\
58 &  01:33:29.4  +30:18:06 & B0253 & 5.50 & $\phn2.72 \pm 0.09\phn$  & $\phn0.050 \pm 0.003\phn$  & $\phn\phn0.017 \pm 0.002\phn\phn$  & $\phn0.688 \pm 0.007\phn$  & $\phn2.05 \pm 0.02\phn$  & $10800 \pm 400\phn\phn$  & $8.115 \pm 0.094$  \\
59 &  01:33:21.2  +30:17:58 & B0254 & 5.56 & $\phn\phn2.2 \pm 0.1\phn\phn$  & $\phn\phn0.33 \pm 0.02\phn\phn$  & $\phn\phn\phn0.03 \pm 0.01\phn\phn\phn$  & $\phn\phn1.20 \pm 0.03\phn\phn$  & $\phn3.73 \pm 0.08\phn$  & $11000 \pm 1000\phn$  & $\phn8.16 \pm 0.19\phn$  \\
60 &  01:32:41.0  +30:24:24 & B0266 & 5.83 & $\phn3.52 \pm 0.05\phn$  & $\phn0.035 \pm 0.002\phn$  & $\phn\phn0.011 \pm 0.001\phn\phn$  & $\phn0.522 \pm 0.003\phn$  & $1.574 \pm 0.008$  & $10300 \pm 300\phn\phn$  & $8.216 \pm 0.084$  \\
61 &  01:32:39.8  +30:22:28 & C395 & 6.12 & $\phn2.31 \pm 0.03\phn$  & $\phn0.039 \pm 0.002\phn$  & $\phn\phn0.013 \pm 0.001\phn\phn$  & $\phn0.614 \pm 0.004\phn$  & $1.839 \pm 0.009$  & $10200 \pm 300\phn\phn$  & $8.129 \pm 0.085$  \\

\enddata

\tablenotetext{a}{Actual position observed based on targets identified
in the Local Group Survey images of \citet{massey-m33m31}.}
\tablenotetext{b}{Closest matching cataloged \ion{H}{2} region for
  observed coordinates.  Catalog abbreviations are: MA = \citet{ma42},
  S = \citet{searle71}, B = \citet{bou74}, C = \citet{cou87}.  Full
  designations of substructure within individual \hii\ regions
  (indicated by the letters at the end of the catalog names) are taken
  from \citet{lgatlas}.}
\tablenotetext{c}{Extinction corrected line fluxes relative to H$\beta=1$.}
\tablenotetext{d}{The same object was observed on multiple slitmasks.}
\end{deluxetable}

\begin{deluxetable}{ccccllllcllll}
\tabletypesize{\scriptsize}
\tablewidth{0pt}
\tablecolumns{13}
\tablecaption{Balmer Series Line Ratios and Derived Properties}
\tablehead{
\colhead{Obj.} & \colhead{EW H$\beta$\tablenotemark{a}} & \colhead{c(H$\beta$)} & 
\colhead{Stellar Absorption EW} & \multicolumn{4}{c}{Observed} & \colhead{} & \multicolumn{4}{c}{Extinction /  Absorption Corrected}\\
\cline{5-8} \cline{10-13}
\colhead{} & \colhead{(\AA)} & \colhead{} & \colhead{(\AA)} &  
\colhead{H$\gamma$} & \colhead{H$\delta$} &
\colhead{H7} & \colhead{H9} & \colhead{} & \colhead{H$\gamma$} & 
\colhead{H$\delta$} & \colhead{H7} & \colhead{H9}
}
\startdata 
1 & $135.3 \pm 0.8\phn\phn$  & $\phn0.46 \pm 0.01\phn$  & $\phn2.0 \pm 0.1\phn$  & 0.390 & 0.175 & 0.106 & 0.032 & &  0.478 & 0.249 & 0.171 & 0.077 \\ 
2 & $\phantom{.}\phn268 \pm 2\phantom{.}\phn\phn\phn$  & $0.155 \pm 0.008$  & $\phn1.6 \pm 0.1\phn$  & 0.436 & 0.2312 & 0.1200 & 0.0552 & &  0.467 & 0.260 & 0.142 & 0.073 \\ 
3 & $\phantom{.}\phn310 \pm 30\phantom{.}\phn\phn$  & $\phn0.29 \pm 0.04\phn$  & $\phantom{.}\phn\phn1 \pm 1\phantom{.}\phn\phn$  & 0.416 & 0.213 & 0.239 & 0.044 & &  0.47 & 0.26 & 0.30 & 0.07 \\ 
4 & $\phantom{.}\phn190 \pm 3\phantom{.}\phn\phn\phn$  & $\phn0.25 \pm 0.01\phn$  & $\phn2.9 \pm 0.2\phn$  & 0.417 & 0.206 & 0.099 & 0.036 & &  0.471 & 0.256 & 0.143 & 0.074 \\ 
5 & $\phantom{.}\phn168 \pm 2\phantom{.}\phn\phn\phn$  & $\phn0.98 \pm 0.06\phn$  & $\phantom{.}\phn\phn7 \pm 1\phantom{.}\phn\phn$  & 0.304 & 0.123 & 0.079 & 0.022 & &  0.47 & 0.26 & 0.20 & 0.10 \\ 
6 & $\phantom{.}\phn351 \pm 3\phantom{.}\phn\phn\phn$  & $0.241 \pm 0.006$  & $\phn1.0 \pm 0.1\phn$  & 0.4303 & 0.2202 & 0.1146 & 0.0566 & &  0.473 & 0.255 & 0.139 & 0.074 \\ 
7 & $\phantom{.}\phn357 \pm 3\phantom{.}\phn\phn\phn$  & $0.266 \pm 0.006$  & $\phn2.2 \pm 0.2\phn$  & 0.4265 & 0.2112 & 0.1248 & 0.0526 & &  0.476 & 0.252 & 0.157 & 0.075 \\ 
8 & $\phantom{.}\phn640 \pm 10\phantom{.}\phn\phn$  & $\phn0.02 \pm 0.05\phn$  & $\phantom{.}\phn\phn8 \pm 2\phantom{.}\phn\phn$  & 0.441 & 0.229 & 0.301 & 0.044 & &  0.47 & 0.26 & 0.34 & 0.07 \\ 
8 & $\phantom{.}\phn440 \pm 20\phantom{.}\phn\phn$  & $\phn0.00 \pm 0.02\phn$  & $\phantom{.}\phn\phn2 \pm 2\phantom{.}\phn\phn$  & 0.468 & 0.255 & 0.325 & 0.066 & &  0.471 & 0.259 & 0.329 & 0.072 \\ 
9\tablenotemark{b} & \nodata & $\phn0.00 \pm 0.09\phn$  & $\phantom{.}\phn11 \pm 2\phantom{.}\phn\phn$  & 0.544 & 0.285 & 0.400 & 0.064 & &  0.52 & 0.23 & 0.34 & 0.004 \\ 
10 & $\phantom{.}\phn193 \pm 6\phantom{.}\phn\phn\phn$  & $\phn0.43 \pm 0.05\phn$  & $\phantom{.}\phn\phn2 \pm 1\phantom{.}\phn\phn$  & 0.391 & 0.193 & 0.120 & 0.041 & &  0.47 & 0.26 & 0.18 & 0.08 \\ 
11 & $110.6 \pm 0.5\phn\phn$  & $\phn0.05 \pm 0.03\phn$  & $\phn1.0 \pm 0.7\phn$  & 0.453 & 0.202 & 0.174 & 0.046 & &  0.47 & 0.22 & 0.19 & 0.06 \\ 
12 & $\phantom{.}\phn750 \pm 70\phantom{.}\phn\phn$  & $\phn0.20 \pm 0.02\phn$  & $\phn2.2 \pm 0.9\phn$  & 0.428 & 0.226 & 0.242 & 0.055 & &  0.466 & 0.260 & 0.283 & 0.072 \\ 
12 & $\phantom{.}4800 \pm 600\phantom{.}\phn$  & $\phn0.15 \pm 0.03\phn$  & $\phantom{.}\phn10 \pm 2\phantom{.}\phn\phn$  & 0.437 & 0.223 & 0.241 & 0.058 & &  0.468 & 0.26 & 0.272 & 0.09 \\ 
13 & $129.5 \pm 0.3\phn\phn$  & $\phn0.05 \pm 0.01\phn$  & $\phn3.9 \pm 0.2\phn$  & 0.426 & 0.194 & 0.098 & 0.026 & &  0.468 & 0.248 & 0.155 & 0.080 \\ 
14 & $\phantom{.}\phn199 \pm 2\phantom{.}\phn\phn\phn$  & $\phn\phn1.0 \pm 0.1\phn\phn$  & $\phantom{.}\phn\phn3 \pm 2\phantom{.}\phn\phn$  & 0.306 & 0.144 & 0.173 & 0.011 & &  0.47 & 0.29 & 0.36 & 0.09 \\ 
15 & $\phantom{.}\phn203 \pm 1\phantom{.}\phn\phn\phn$  & $0.220 \pm 0.008$  & $\phn0.7 \pm 0.1\phn$  & 0.435 & 0.2207 & 0.1261 & 0.0581 & &  0.474 & 0.253 & 0.150 & 0.075 \\ 
16 & $116.0 \pm 0.4\phn\phn$  & $\phn0.00 \pm 0.02\phn$  & $\phn2.5 \pm 0.2\phn$  & 0.456 & 0.222 & 0.283 & 0.042 & &  0.479 & 0.254 & 0.316 & 0.083 \\ 
17 & $54.46 \pm 0.08\phn$  & $\phn0.00 \pm 0.01\phn$  & $3.80 \pm 0.04$  & 0.406 & 0.162 & 0.068 & 0.00008 & &  0.474 & 0.256 & 0.173 & 0.114 \\ 
18 & $\phantom{.}\phn442 \pm 1\phantom{.}\phn\phn\phn$  & $\phn0.09 \pm 0.01\phn$  & $\phn1.6 \pm 0.3\phn$  & 0.452 & 0.238 & 0.156 & 0.063 & &  0.471 & 0.256 & 0.172 & 0.074 \\ 
19 & $\phantom{.}\phn240 \pm 10\phantom{.}\phn\phn$  & $\phn\phn0.5 \pm 0.1\phn\phn$  & $\phantom{.}\phn\phn1 \pm 3\phantom{.}\phn\phn$  & 0.382 & 0.142 & 0.082 & 0.014 & &  0.47 & 0.20 & 0.13 & 0.03 \\ 
20 & $248.3 \pm 0.8\phn\phn$  & $\phn0.30 \pm 0.02\phn$  & $\phn0.9 \pm 0.6\phn$  & 0.417 & 0.218 & 0.228 & 0.055 & &  0.467 & 0.260 & 0.278 & 0.073 \\ 
21 & $170.6 \pm 0.3\phn\phn$  & $0.095 \pm 0.004$  & $2.24 \pm 0.04$  & 0.4502 & 0.2154 & 0.1294 & 0.0423 & &  0.481 & 0.249 & 0.163 & 0.076 \\ 
22 & $\phantom{.}\phn256 \pm 9\phantom{.}\phn\phn\phn$  & $\phn0.13 \pm 0.02\phn$  & $\phn0.4 \pm 0.3\phn$  & 0.446 & 0.239 & 0.131 & 0.064 & &  0.468 & 0.259 & 0.145 & 0.073 \\ 
23 & $\phantom{.}\phn744 \pm 3\phantom{.}\phn\phn\phn$  & $0.270 \pm 0.006$  & $\phn0.0 \pm 0.3\phn$  & 0.426 & 0.2225 & 0.1401 & 0.0622 & &  0.470 & 0.257 & 0.165 & 0.075 \\ 
24 & $\phantom{.}\phn350 \pm 10\phantom{.}\phn\phn$  & $\phn0.34 \pm 0.01\phn$  & $\phn0.0 \pm 0.2\phn$  & 0.424 & 0.209 & 0.113 & 0.061 & &  0.478 & 0.249 & 0.139 & 0.077 \\ 
25 & $173.4 \pm 0.1\phn\phn$  & $0.288 \pm 0.007$  & $0.80 \pm 0.08$  & 0.4225 & 0.2117 & 0.1258 & 0.0528 & &  0.474 & 0.254 & 0.159 & 0.075 \\ 
26 & $179.7 \pm 0.2\phn\phn$  & $\phn0.14 \pm 0.01\phn$  & $\phn1.9 \pm 0.1\phn$  & 0.436 & 0.219 & 0.1083 & 0.0438 & &  0.473 & 0.255 & 0.140 & 0.074 \\ 
27 & $\phantom{.}\phn335 \pm 4\phantom{.}\phn\phn\phn$  & $0.150 \pm 0.008$  & $\phn0.7 \pm 0.2\phn$  & 0.446 & 0.2328 & 0.1450 & 0.0631 & &  0.473 & 0.255 & 0.162 & 0.074 \\ 
28 & $\phantom{.}\phn435 \pm 8\phantom{.}\phn\phn\phn$  & $\phn0.08 \pm 0.01\phn$  & $\phn4.5 \pm 0.4\phn$  & 0.446 & 0.233 & 0.129 & 0.051 & &  0.469 & 0.259 & 0.154 & 0.073 \\ 
29 & $125.9 \pm 0.6\phn\phn$  & $\phn0.00 \pm 0.02\phn$  & $\phn3.1 \pm 0.3\phn$  & 0.450 & 0.222 & 0.293 & 0.046 & &  0.475 & 0.255 & 0.33 & 0.088 \\ 
30 & $\phantom{.}\phn101 \pm 1\phantom{.}\phn\phn\phn$  & $\phn0.08 \pm 0.05\phn$  & $\phantom{.}\phn\phn1 \pm 1\phantom{.}\phn\phn$  & 0.448 & 0.205 & 0.186 & 0.048 & &  0.47 & 0.22 & 0.21 & 0.06 \\ 
31 & $\phantom{.}\phn191 \pm 3\phantom{.}\phn\phn\phn$  & $\phn0.03 \pm 0.01\phn$  & $\phn2.5 \pm 0.2\phn$  & 0.452 & 0.233 & 0.145 & 0.046 & &  0.470 & 0.257 & 0.171 & 0.074 \\ 
32 & $\phantom{.}\phn250 \pm 3\phantom{.}\phn\phn\phn$  & $0.211 \pm 0.009$  & $\phn2.6 \pm 0.1\phn$  & 0.428 & 0.2123 & 0.1185 & 0.0463 & &  0.473 & 0.254 & 0.155 & 0.075 \\ 
33 & $\phantom{.}1300 \pm 200\phantom{.}\phn$  & $\phn0.31 \pm 0.02\phn$  & $\phn3.6 \pm 0.8\phn$  & 0.420 & 0.208 & 0.130 & 0.052 & &  0.475 & 0.253 & 0.167 & 0.075 \\ 
34 & $\phantom{.}\phn820 \pm 70\phantom{.}\phn\phn$  & $\phn0.02 \pm 0.06\phn$  & $\phantom{.}\phn\phn1 \pm 5\phantom{.}\phn\phn$  & 0.462 & 0.222 & 0.180 & 0.073 & &  0.47 & 0.23 & 0.19 & 0.08 \\ 
35 & $\phantom{.}\phn237 \pm 1\phantom{.}\phn\phn\phn$  & $\phn0.09 \pm 0.02\phn$  & $\phn0.3 \pm 0.3\phn$  & 0.452 & 0.245 & 0.162 & 0.067 & &  0.468 & 0.259 & 0.173 & 0.073 \\ 
36 & $146.1 \pm 0.7\phn\phn$  & $\phn0.15 \pm 0.04\phn$  & $\phn2.2 \pm 0.8\phn$  & 0.428 & 0.218 & 0.117 & 0.045 & &  0.47 & 0.26 & 0.16 & 0.08 \\ 
37 & $\phn83.2 \pm 0.2\phn\phn$  & $\phn0.00 \pm 0.02\phn$  & $\phn3.3 \pm 0.2\phn$  & 0.444 & 0.203 & 0.115 & 0.041 & &  0.479 & 0.253 & 0.175 & 0.106 \\ 
38 & $\phantom{.}\phn600 \pm 10\phantom{.}\phn\phn$  & $\phn0.90 \pm 0.03\phn$  & $\phn0.9 \pm 4\phantom{.}\phn\phn$  & 0.338 & 0.160 & 0.110 & 0.032 & &  0.47 & 0.26 & 0.19 & 0.06 \\ 
39 & $\phantom{.}\phn162 \pm 3\phantom{.}\phn\phn\phn$  & $\phn0.00 \pm 0.02\phn$  & $\phn5.7 \pm 0.2\phn$  & 0.439 & 0.202 & 0.128 & 0.031 & &  0.479 & 0.254 & 0.185 & 0.092 \\ 
40 & $\phantom{.}\phn367 \pm 8\phantom{.}\phn\phn\phn$  & $\phn0.11 \pm 0.01\phn$  & $\phn4.4 \pm 0.3\phn$  & 0.443 & 0.220 & 0.145 & 0.046 & &  0.476 & 0.252 & 0.178 & 0.075 \\ 
41 & $\phantom{.}\phn280 \pm 30\phantom{.}\phn\phn$  & $\phn0.26 \pm 0.04\phn$  & $\phantom{.}\phn\phn1 \pm 1\phantom{.}\phn\phn$  & 0.421 & 0.205 & 0.115 & 0.025 & &  0.47 & 0.24 & 0.14 & 0.04 \\ 
42 & $12.03 \pm 0.07\phn$  & $\phn\phn0.5 \pm 0.1\phn\phn$  & $\phn2.5 \pm 0.2\phn$  & 0.18 & 0.04 & -0.27 & -0.33 & &  0.47 & 0.40 & 0.09 & 0.0003 \\ 
43 & $\phantom{.}\phn218 \pm 7\phantom{.}\phn\phn\phn$  & $\phn0.10 \pm 0.02\phn$  & $\phn2.0 \pm 0.2\phn$  & 0.448 & 0.228 & 0.140 & 0.052 & &  0.474 & 0.254 & 0.166 & 0.075 \\ 
44 & $\phn51.6 \pm 0.2\phn\phn$  & $\phn0.00 \pm 0.03\phn$  & $2.56 \pm 0.06$  & 0.437 & 0.182 & 0.104 & 0.006 & &  0.485 & 0.251 & 0.183 & 0.097 \\ 
45 & $\phantom{.}\phn117 \pm 4\phantom{.}\phn\phn\phn$  & $\phn0.00 \pm 0.03\phn$  & $\phn4.3 \pm 0.3\phn$  & 0.445 & 0.191 & 0.117 & 0.006 & &  0.49 & 0.250 & 0.179 & 0.083 \\ 
46 & $\phantom{.}\phn137 \pm 3\phantom{.}\phn\phn\phn$  & $\phn0.00 \pm 0.02\phn$  & $\phn7.0 \pm 0.4\phn$  & 0.425 & 0.193 & 0.244 & 0.037 & &  0.469 & 0.259 & 0.301 & 0.109 \\ 
47 & $\phantom{.}\phn250 \pm 2\phantom{.}\phn\phn\phn$  & $0.054 \pm 0.008$  & $\phn1.6 \pm 0.1\phn$  & 0.456 & 0.2369 & 0.1896 & 0.0558 & &  0.473 & 0.255 & 0.209 & 0.074 \\ 
48 & $216.2 \pm 0.2\phn\phn$  & $0.155 \pm 0.007$  & $\phn2.2 \pm 0.1\phn$  & 0.4395 & 0.2167 & 0.1646 & 0.0490 & &  0.476 & 0.252 & 0.200 & 0.075 \\ 
49 & $\phantom{.}\phn119 \pm 4\phantom{.}\phn\phn\phn$  & $\phn0.29 \pm 0.06\phn$  & $\phn2.4 \pm 0.8\phn$  & 0.401 & 0.195 & 0.136 & 0.032 & &  0.47 & 0.26 & 0.20 & 0.08 \\ 
50 & $\phn27.6 \pm 0.4\phn\phn$  & $\phn0.25 \pm 0.05\phn$  & $\phn1.5 \pm 0.2\phn$  & 0.373 & 0.124 & 0.000 & -0.045 & &  0.47 & 0.23 & 0.10 & 0.06 \\ 
51 & $214.9 \pm 0.3\phn\phn$  & $0.273 \pm 0.009$  & $\phn3.0 \pm 0.1\phn$  & 0.420 & 0.1923 & 0.1659 & 0.0384 & &  0.481 & 0.246 & 0.222 & 0.077 \\ 
52 & $\phantom{.}\phn138 \pm 2\phantom{.}\phn\phn\phn$  & $\phn0.07 \pm 0.02\phn$  & $\phn2.7 \pm 0.2\phn$  & 0.445 & 0.213 & 0.094 & 0.040 & &  0.478 & 0.249 & 0.131 & 0.077 \\ 
53 & $\phantom{.}\phn121 \pm 1\phantom{.}\phn\phn\phn$  & $\phn0.08 \pm 0.02\phn$  & $\phn6.3 \pm 0.2\phn$  & 0.417 & 0.167 & 0.074 & 0.028 & &  0.478 & 0.247 & 0.161 & 0.085 \\ 
54 & $\phantom{.}\phn250 \pm 10\phantom{.}\phn\phn$  & $\phn0.18 \pm 0.02\phn$  & $\phn3.2 \pm 0.4\phn$  & 0.432 & 0.211 & 0.101 & 0.044 & &  0.475 & 0.252 & 0.136 & 0.076 \\ 
55 & $\phantom{.}\phn455 \pm 3\phantom{.}\phn\phn\phn$  & $0.082 \pm 0.005$  & $1.83 \pm 0.07$  & 0.4539 & 0.2316 & 0.2481 & 0.0558 & &  0.476 & 0.253 & 0.276 & 0.074 \\ 
56 & $\phantom{.}3100 \pm 500\phantom{.}\phn$  & $0.418 \pm 0.009$  & $\phantom{.}\phn10 \pm 10\phantom{.}\phn$  & 0.421 & 0.184 & 0.156 & 0.0483 & &  0.50 & 0.24 & 0.21 & 0.07 \\ 
57 & $243.3 \pm 0.7\phn\phn$  & $\phn0.18 \pm 0.01\phn$  & $\phn0.4 \pm 0.2\phn$  & 0.439 & 0.231 & 0.173 & 0.062 & &  0.470 & 0.257 & 0.196 & 0.074 \\ 
58 & $117.3 \pm 0.3\phn\phn$  & $\phn0.09 \pm 0.04\phn$  & $\phn5.3 \pm 0.7\phn$  & 0.415 & 0.194 & 0.120 & 0.034 & &  0.47 & 0.26 & 0.19 & 0.11 \\ 
59 & $20.76 \pm 0.04\phn$  & $\phn0.22 \pm 0.07\phn$  & $\phn1.1 \pm 0.2\phn$  & 0.37 & 0.088 & 0.00 & -0.08 & &  0.47 & 0.19 & 0.11 & 0.04 \\ 
60 & $\phantom{.}\phn164 \pm 4\phantom{.}\phn\phn\phn$  & $\phn0.23 \pm 0.02\phn$  & $\phn2.4 \pm 0.2\phn$  & 0.422 & 0.202 & 0.105 & 0.039 & &  0.475 & 0.252 & 0.148 & 0.075 \\ 
61 & $\phantom{.}\phn\phn95 \pm 1\phantom{.}\phn\phn\phn$  & $\phn0.00 \pm 0.01\phn$  & $\phn4.9 \pm 0.1\phn$  & 0.420 & 0.186 & 0.107 & 0.024 & &  0.473 & 0.257 & 0.188 & 0.112 \\ 

\enddata
\tablecomments{\ Line fluxes are given relative to H$\beta$=1 and rounded
  to the precision of one significant figure in the error.}
\tablenotetext{a}{The equivalent width of H$\beta$ in emission.}
\tablenotetext{b}{Because there is no discernible continuum emission
  associated with the nebular line emission from this \hii\ region, the
  equivalent width of the H$\beta$ emission is undefined.}
\label{hlinetable}
\end{deluxetable}
\clearpage
\end{landscape}

\end{document}